\documentclass[12pt]{article}
\input epsf
\ifx\epsffile\undefined
\newlength{\epsfysize}
\def\epsffile#1#2#3#4]#5{}
\fi

\newcommand{\dr}{{\footnotesize{$\overline{{\rm DR}}$}} }
\newcommand{\ms}{{\footnotesize{$\overline{{\rm MS}}$}} }
\def\gsim{\raise.3ex\hbox{$>$\kern-.75em\lower1ex\hbox{$\sim$}}}
\def\lsim{\raise.3ex\hbox{$<$\kern-.75em\lower1ex\hbox{$\sim$}}}

\begin{document}

\begin{flushright} {\small SLAC--PUB--7403} \end{flushright}

\vspace{2cm}

\baselineskip=24pt
\begin{center} \large
The Strong Coupling Constant in Grand Unified Theories\footnote{Work
supported by Department of Energy Contract DE--AC03--76SF00515.}
\end{center}

\baselineskip=32pt

\centerline{Damien M. Pierce}

\baselineskip=22pt

\begin{center}
\footnotesize\it
Stanford Linear Accelerator Center\\
\baselineskip=13pt
Stanford University\\
Stanford, California 94309
\end{center}

\vspace{1cm}

\begin{abstract}
The prediction of the strong coupling constant in grand unified
theories is reviewed, first in the standard model, then in the
supersymmetric version. Various corrections are considered. The
predictions in both supergravity-induced and gauge-mediated
supersymmetry breaking models are discussed. In the region of
parameter space without large fine tuning the strong coupling is
predicted to be $\alpha_s(M_Z)\,\gsim\,0.13$. Imposing
$\alpha_s(M_Z)=0.118$, we require a unification scale threshold
correction of typically $-2\%$, which is accommodated by some GUT
models but in conflict with others.
\end{abstract}

\vspace{2cm}

\centerline{\sl Submitted to the proceedings of the 1996 SLAC Summer
Institute}

\vfill

\pagebreak

\normalsize\baselineskip=15pt

\section{\large Introduction}

In the standard model, given values of the three gauge couplings, the
Yukawa couplings, the Higgs boson self-coupling, and a dimensionful
observable (e.g. the muon lifetime, a gauge boson mass or a quark
mass), any other observable can then be computed. In a grand-unified
theory (GUT), one need only input two of the gauge couplings, then the
third one can be predicted, as well as other observables.

Thus GUT theories are more predictive. Because the SU(2) and U(1)
couplings are measured quite precisely, we will examine the prediction
of the strong coupling constant in grand-unified theories, both in
the standard model and in supersymmetric models. We start with the
one-loop results and then proceed to discuss the next higher-order
corrections. The inclusion of these corrections leads to a precise
prediction of the strong coupling constant.

\section{\large Renormalization Group Equations}

We are interested in measuring gauge couplings at the weak-scale or
below, and running the gauge couplings up to higher scales. The
solution of the renormalization group equations (RGE's) accurately
describes the evolution of the couplings, even as they are evolved
over many decades. At one-loop, the renormalization group equations
for the gauge couplings are
\begin{equation}
{dg_i\over dt} = {b_i\over 16\pi^2}g_i^3,\qquad t=\ln{Q\over Q_0}\ ,
\end{equation}
where the $b_i$ are the one-loop beta-constants. These constants receive
contributions from every particle which circulates in the one-loop
gauge-boson self-energy diagram. For example, for a fermion doublet
$(\nu,e)$, $\Delta b_2=1/3$.

The one-loop renormalization group equation is easily solved. The inverse
of the gauge coupling evolves linearly with the log of the scale,
\begin{equation}
\alpha_i^{-1}(Q) = \alpha_i^{-1}(Q_0) - {b_i\over2\pi}\ln{Q\over Q_0}\ .
\end{equation}
If we assume that the couplings do unify at some scale, we have that
\begin{equation}
{b_1-b_2\over\alpha_3(Q)} + {b_2-b_3\over\alpha_1(Q)} +
{b_3-b_1\over\alpha_2(Q)} = 0\ .\label{alpha}
\end{equation}
This equation is valid for any scale $Q$ between the weak scale and
the unification scale.  Next, we deduce the \ms values of the U(1) and
SU(2) gauge couplings from the quantities $\hat\alpha^{-1} =
127.90\pm0.09$ and $\hat s^2=0.2315\pm0.0004$ \cite{PDG} ($\hat\alpha$
and $\hat s^2$ are the \ms values of the electromagnetic coupling and
sine-squared weak mixing angle evaluated at $M_Z$)
\begin{equation}
\alpha_1^{-1}(M_Z)=58.97\pm0.05\ ,\qquad
\alpha_2^{-1}(M_Z)=29.61\pm0.05~.\label{g1g2}
\end{equation}
Given the standard model values of the one-loop beta-constants
\begin{equation}
b_1={41\over10}\ ,\qquad b_2=-{19\over6}\ ,\qquad b_3=-7\ ,
\end{equation}
we determine the prediction of the strong coupling constant in the
standard model
\begin{equation}
\alpha_s(M_Z)=0.07 \qquad\quad \mbox{(standard model, one-loop)}
\end{equation}
with negligible error.  Comparing with the measured value as quoted by
the Particle Data Group \cite{PDG}, $\alpha_s(M_Z)=0.118\pm0.003$, we
see that the standard model prediction of the strong coupling
constant is about 16 standard deviations too small.  Including higher
order corrections cannot help repair this situation by more than a few
standard deviations. In order to remedy this huge discrepancy, we need
to go beyond the standard model, adding matter to change the
beta-constants $b_i$. We also need to specify the mass scale of the
new matter.

Rather than determining what additional matter content will lead to
successful gauge coupling unification in an ad hoc fashion, we will
examine the implications of a motivated model. It is widely accepted
that the standard model is an effective theory. Some new physics must
become manifest at scales below the multi-TeV scale. A promising model
of new physics is supersymmetry \cite{review}, which, among other
virtues, explains how a theory with a weak-scale elementary Higgs
boson is stable with respect to the Planck scale.  Supersymmetry also
naturally explains the breaking of electroweak symmetry \cite{ewsb},
and it typically contains a natural dark matter candidate \cite{dm}.

In order to make the standard model supersymmetric, we are required to
add a specific set of new particles, the superpartners. For every
standard-model gauge boson we must add a spin-1/2 majorana particle, a
gaugino.  And for every standard-model fermion we must add a spin-0
partner, a squark or a slepton. In order to give both up- and
down-type fermions a mass, and to ensure anomaly cancellation, we must
also add an additional Higgs doublet. We refer to the two Higgs
doublets as ``up-type'' and ``down-type''. This additional
particle content defines the minimal supersymmetric standard
model (MSSM). In the simplest and most often considered versions of
the MSSM with high-energy inputs, the entire supersymmetric spectrum
scales with one or two parameters which must be near the weak scale by
naturalness considerations.

Given this new matter content and the superpartner mass scale, which
we take to be $M_Z$ for the moment, we can take a first order look at
the prediction of the strong coupling constant in the MSSM. As before
we use Eq.~(\ref{alpha}) to solve for $\alpha_s(M_Z)$ as a function of
$\alpha_1(M_Z)$ and $\alpha_2(M_Z)$. Now the beta-constants are
different.  Because of the new matter content, we have the following
changes from the standard-model values
\begin{equation}
b_1=4.1\rightarrow6.6\ ,\qquad b_2=-3.2\rightarrow1\ ,\qquad
b_3=-7\rightarrow-3\ .
\end{equation}
We use the central values and errors in Eq.~(\ref{g1g2}) to
find
\begin{equation}
\alpha_s(M_Z)=0.116\pm0.001\qquad\quad\mbox{(MSSM, one-loop)}
\label{pred}
\end{equation}
The new matter content mandated by supersymmetry brings the prediction
for the strong coupling constant to the measured value
($0.118\pm0.003$) within one standard deviation! This could be a
coincidence, or it could be a hint that we are on the right track by
considering minimal low energy supersymmetry. Next we will consider
corrections to the prediction of $\alpha_s$ in the MSSM.

\section{\large Corrections to the prediction of $\alpha_s$}

In this section we consider two sources of corrections to the
prediction of the strong coupling constant in the MSSM. The first
involves improving the evolution of the couplings from their initial
values at the grand unification scale to the weak scale (or, the
supersymmetry breaking scale). We simply extend the renormalization
group equations to one higher loop order. This means that in addition
to resumming logarithms of the form $[\alpha/4\pi\ln(M_{\rm GUT}/
M_Z)]^n$ from the one-loop RGE, we now resum logarithms of the form
$[(\alpha/4\pi)^2 \ln(M_{\rm GUT}/M_Z)]^n$ from the two-loop terms.
At two-loops it is necessary to run both the gauge and Yukawa
couplings together, as they form a coupled set of differential
equations. The general form for either a gauge or Yukawa coupling RGE
at two-loops is \cite{RGE}
\begin{equation}
{dg_i\over dt} = g_i\biggl\{{b_{ij}\over16\pi^2}g_j^2 + 
{b_{ijk}\over(16\pi^2)^2}g_j^2g_k^2\biggr\}\ .
\end{equation}
Since we can safely ignore the small Yukawa couplings of the first two
generations, the set $g_i$ includes $\{g_1,~g_2,~g_3,~\lambda_t,
~\lambda_b,~\lambda_\tau\}$. The equations are readily solved
numerically. The correction due to the two-loop RGE's is large and
positive. The prediction of $\alpha_s(M_Z)$ increases by about
$0.01$.

The second source of corrections we will consider in this section are
the supersymmetric threshold corrections. These are divided into two
types, the logarithmic corrections and the finite
(i.e. non-logarithmic) corrections.  We will discuss these in turn.

When we arrived at the prediction (\ref{pred}) we assumed that the
masses of the superpartners were equal to $M_Z$. However, we know from
collider searches that many of the superpartner masses must be heavier
than $M_Z$, and in general they could be an order of magnitude heavier
without straining naturalness considerations too much. As we evolve
the gauge couplings down from high scales, we must decouple each
superpartner contribution in turn as we cross the mass threshold.
Hence we arrive at logarithmic corrections of the form $\Delta
b_i\ln(M_{\rm susy}/M_Z)$. From these corrections we determine the
shift in the prediction of the strong coupling and find, in general,
for a particle of mass $M>M_Z$ with $\Delta b_i$ contributions to the
beta-constants
\begin{equation}
\Delta\alpha_s^{-1}(M_Z) = -{1\over2\pi(b_1-b_2)}\biggl[
(b_2-b_3)\Delta b_1+(b_3-b_1)\Delta b_2+(b_1-b_2)\Delta b_3
\biggr]\ln{M\over M_Z}.\label{mas}
\end{equation}
Note that the beta-constants $b_i$ in this equation {\em include} the
entire superpartner spectrum.  Plugging in all the superpartners, we
arrive at the following supersymmetric threshold correction
\begin{eqnarray}
\Delta\alpha_s^{-1}(M_Z)&=&{1\over28\pi}\biggl[
3\ln{M_L^3M_Q^7\over M_D^3M_E^2M_U^5} + 32\ln{M_2\over M_Z} 
- 28\ln{M_3\over M_Z}\label{dals}\\
&& \qquad +~~3\ln{M_H\over M_Z}~
+ 12\ln{|\mu|\over M_Z}\biggr]\ .\nonumber
\end{eqnarray}
In the first term $M_Q$ and $M_L$ ($M_U$, $M_D$, and $M_E$) denote the
left-handed (right-handed) squark and slepton masses. This term is
easily modified to account for generation-dependent masses.

There are a few aspects of this equation worth pointing out. First,
notice that the first term does not contain $M_Z$. This is because the
squarks and sleptons are in complete SU(5) multiplets, and degenerate
complete multiplets do not affect $\alpha_s$ at one loop. It happens
that this particular combination of soft squark and slepton masses is
near unity in the entire parameter space of both of the models that we
consider in the following section. Hence, the squarks and sleptons do
not significantly affect the prediction of $\alpha_s(M_Z)$.

Another point is that the SU(2) and SU(3) gauginos contribute
corrections which largely combine into the term $(1/\pi)\ln M_2/M_3$.
In GUT models the gaugino masses are degenerate above the unification
scale.  At one-loop they are renormalized proportional to the
corresponding gauge couplings, so the ratio $M_2/M_3=\alpha_2/
\alpha_3$. At the weak scale, this ratio is about
$8/30\simeq0.27$. Hence, this term contributes about $+0.006$ to
$\alpha_s(M_Z)$, independent of parameter space.

The remaining terms include the residual term from the SU(2) gaugino
mass, and the contributions of the heavy Higgs bosons and Higgsinos.
If we take these three masses to be characterized by a single scale
$M_{\rm susy}$, then these terms combine to give
\begin{equation}
\Delta\alpha_s^{-1}(M_Z)={19\over28\pi}\ln {M_{\rm susy}\over M_Z}~.
\label{msusy}
\end{equation}
Hence, as we increase the supersymmetric mass scale the prediction of
the strong coupling constant decreases. This makes sense, since, in
the limit that the entire supersymmetric spectrum is raised above
$M_{\rm GUT}$ we should recover the prediction of the standard model.
(In fact $(0.116)^{-1} + (19/28\pi)\ln M_{\rm GUT}/M_Z =
(0.064)^{-1}$.)  If $M_{\rm susy}=1$ TeV this correction yields
$\Delta\alpha_s(M_Z)=-0.007$. Note that the largest contribution to
Eq.~(\ref{msusy}) is due to the $\ln(|\mu|/M_Z)$ term. Because we
impose electroweak symmetry breaking, $|\mu|/M_Z$ is a measure of the
fine tuning necessary to obtain the $Z$-mass from the input mass
parameters. The predicted strong coupling is larger in the region of
no fine tuning ($|\mu|/M_Z\,\lsim\,2$) than in the region of
appreciable fine tuning ($|\mu|/M_Z\,\gsim\,10$).

There is one last point about Eq.~(\ref{dals}). Because of the
particle content of the supersymmetric standard model, the particles
which are charged under SU(2) always come into the expression for
$\Delta\alpha_s^{-1}$ with a positive coefficient in front of the
logarithm of the mass. The SU(2) singlets always come with a
negative coefficient. Hence, heavy SU(2) doublets, for example $\tilde
W$, $H$ or $\tilde H$, will decrease $\alpha_s$, and heavy SU(2)
singlets, e.g. the gluino, will increase the prediction of the strong
coupling.

Besides the logarithmic corrections of Eq.~(\ref{dals}), there are
also finite corrections. These arise when the full one-loop correction
to $\hat s^2$ is taken into account. Taking the electromagnetic
constant, the $Z$-boson mass, and the muon lifetime as inputs, we
determine the \dr \cite{dr} renormalized weak mixing angle
\cite{dfs},
\begin{equation}
\hat s^2 \hat c^2 = {\pi\hat\alpha\over\sqrt2 G_\mu M_Z^2(1-\Delta\hat
r)}\ , \qquad \Delta\hat r={\hat\Pi_{WW}(0)\over M_W^2} -
{\hat\Pi_{ZZ}(M_Z^2)\over M_Z^2} + \delta_{\rm vb}\ .
\end{equation}
The correction $\Delta\hat r$ is comprised of the real and transverse
\dr gauge-boson self-energy contribution (the
oblique correction), and the vertex and box diagram contributions,
$\delta_{\rm vb}$ (the non-universal part) \cite{pbmz}. The oblique
correction contains both logarithmic and finite contributions, while
$\delta_{\rm vb}$ is purely finite. The finite contributions decouple
as $M_Z^2/M_{\rm susy}^2$ for large supersymmetric masses.

In the regions of parameter space where some SU(2) non-singlet
particles are of order $M_Z$, the finite corrections to the weak
mixing angle can substantially increase the prediction of the strong
coupling constant \cite{cpp,bmp}. We illustrate this in Fig.~1.  We
show the prediction of the strong coupling with and without taking the
finite corrections into account in the supergravity model described
below.  If all the supersymmetric particles are above a few hundred
GeV, the finite corrections are negligible. In that case the
low-energy threshold corrections are well approximated by the
logarithms in Eq.~(\ref{dals}).

\begin{figure}[t]
\epsfysize=2.5in
\epsffile[60 500 160 720]{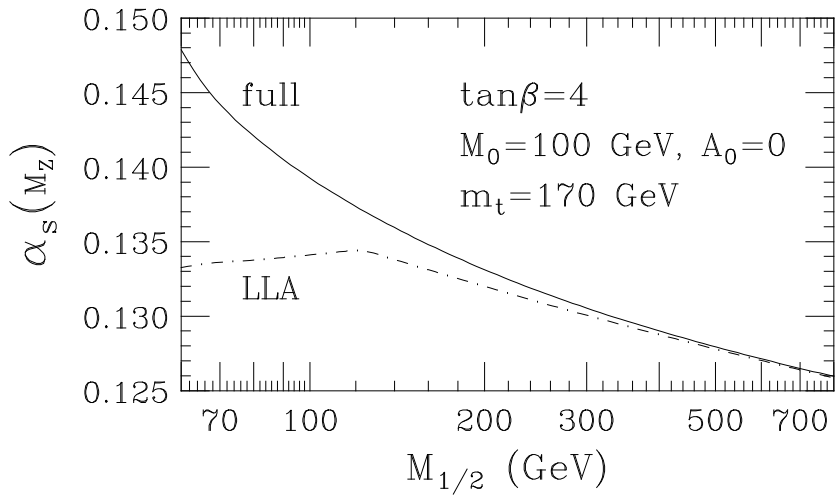}
\begin{center}
\parbox{5.5in}{
\caption[]{\small The prediction of $\alpha_s(M_Z)$ with (solid) and
without (dot-dashed) including the finite corrections.  From
Ref.~\cite{bmp}.}}
\end{center}
\end{figure}

\section{\large $\alpha_s(M_Z)$ at two-loops in two supersymmetric models}

Utilizing the two-loop renormalization group equations and the
weak-scale threshold corrections we are now in a position to present
the improved prediction of the strong coupling constant in the
supersymmetric case. We still need to specify the supersymmetric
particle spectrum. In what follows we consider two models in turn,
first a minimal supergravity model, then a simple model with
gauge-mediated supersymmetry breaking.

\subsection{Minimal supergravity model}

In the minimal supergravity model there are universal
soft-supersymmetry breaking parameters induced at the grand
unification scale. These include a gaugino mass $M_{1/2}$, a scalar
mass $M_0$, and a trilinear scalar coupling $A_0$. Also, we require
that electroweak symmetry is broken radiatively \cite{ewsb}. This
happens naturally, as the large top-quark Yukawa coupling drives the
up-type Higgs mass-squared negative near the weak scale. Given the
$Z$-boson mass and $\tan\beta$ (the ratio of up-type Higgs boson
vacuum expectation value (vev) to down-type), we impose electroweak
symmetry breaking and this allows us to solve for the masses of the
heavy Higgs bosons and their superpartners, $M_H$ and $|\mu|$.

In Fig.~2 we show the prediction of the strong coupling in the
$M_0$, $M_{1/2}$ plane. After including the corrections the
supersymmetric prediction is not as consistent with the measured value
$0.118\pm0.003$. We find that for squark masses below about 1
TeV (where the model is more natural) $\alpha_s$ is predicted to be
greater than 0.127. For the smallest supersymmetric masses we find
numbers larger than 0.140. The predicted value is three to seven
standard deviations larger than the measured value.

\begin{figure}[t]
\epsfysize=2.5in
\epsffile[-45 420 55 720]{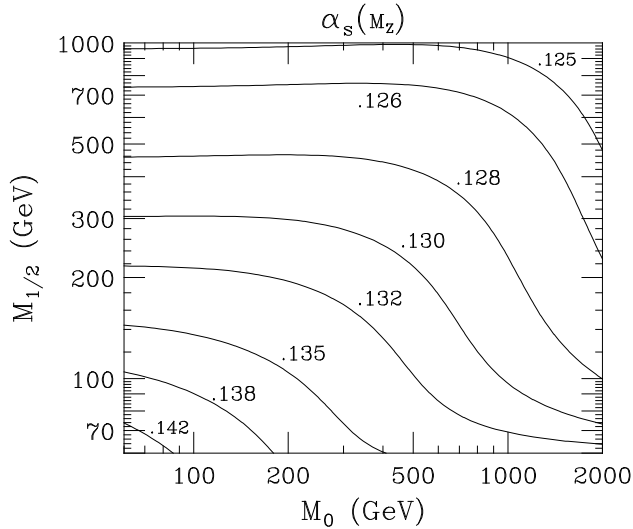}
\begin{center}
\parbox{5.5in}{
\caption[]{\small The prediction of $\alpha_s(M_Z)$ 
in the $M_0$, $M_{1/2}$ plane, with $\tan\beta=2$ and $A_0=0$.
From Ref.~\cite{bmp}.}}
\end{center}
\end{figure}

The prediction of $\alpha_s$ depends slightly on $m_t$. It changes by
about 0.001 if $m_t$ changes by 10 GeV. Of the three input parameters
$M_Z,\ G_\mu,$ and $\hat\alpha$, only $\hat\alpha$ has an appreciable
error. Changing $\hat\alpha$ by 1-$\sigma$ changes $\alpha_s$ by about
0.001. The prediction weakly depends on $\tan\beta$ because at small
($\sim1$) or large ($\gsim30$) $\tan\beta$ the top, bottom, and/or
tau-Yukawa couplings become large, and they enter into the two-loop
renormalization group equations of the gauge couplings. The prediction
for $\alpha_s(M_Z)$ can be lowered by about 0.002 at the extreme
values of $\tan\beta$\footnote{The change $-0.002$ is due solely to
the Yukawa couplings entering into the two-loop RGE's. There may be
an additional dependence because the particle spectrum depends on
$\tan\beta$.}.  The strong coupling is also insensitive to the sign of
the Higgsino mass term $\mu$. In Fig.~2 and the following figures we
set $\mu>0$.

Hence we find that we cannot avoid the large values of $\alpha_s$
shown in Fig.~2. We will give an interpretation of these large
numbers after discussing the gauge-mediated case.

\subsection{Minimal gauge-mediated model}

Models with gauge-mediated supersymmetry breaking are an attractive
alternative to the supergravity models. In the supergravity model we
considered, we chose universal boundary conditions, which suppress
dangerous squark- and slepton-mediated flavor changing neutral
currents. However, there is no symmetry which protects this choice of
boundary conditions, and hence it is an artificial imposition.  The
advantage of the gauge-mediated models is that flavor changing neutral
currents are automatically suppressed, since the soft supersymmetry
breaking masses which are induced by the gauge interactions are flavor
diagonal and generation independent.

In the simplest models with gauge-mediated supersymmetry breaking
\cite{dnns}, there is a messenger sector with the superpotential
interaction
\begin{equation}
W=\lambda S M \overline{M}\label{W}\ ,
\end{equation}
where $S$ is a standard-model singlet superfield, $M$ and $\overline
M$ are a pair of messenger fields which are vector-like under the
standard-model gauge group, and $\lambda$ is the messenger Yukawa
coupling.  In a grand unified theory, $M$ and $\overline M$ come in
full SU(5) representations. We will consider $n_5$ 5+$\overline{5}$
pairs and $n_{10}$ 10+$\overline{10}$ representations. Perturbative
unification of the gauge couplings is ensured if we allow at most
($n_5,n_{10}$) = (1,1) or (4,0). 

In these models the singlet superfield S couples to a sector in which
supersymmetry is dynamically broken, and as a result the it acquires
both a vev ($S$) and an F-term ($F$). This in turn generates
supersymmetry-conserving diagonal entries and supersymmetry-violating
off diagonal entries in the $M,\ \overline M$ scalar mass matrix.  The
$M$ and $\overline M$ fields enter into loop diagrams with standard
model fields on the external legs, thereby generating soft
supersymmetry breaking masses for the superpartners and the Higgs
bosons. The gaugino and scalar masses are generated at one- and
two-loops, respectively, and in the limit $F\ll\lambda S^2$ are given
by \cite{dnns}
\begin{eqnarray}
M_i(M) &=& (n_5+3n_{10}){\alpha_i(M)\over4\pi}\Lambda\ ,\label{bc1}\\
\tilde m^2(M) &=& 2(n_5+3n_{10})\sum_{i=1}^3C_i
\left({\alpha_i(M)\over4\pi}\right)^2\Lambda^2\ ,\label{bc2}
\end{eqnarray}
where $M=\lambda S$ is the messenger mass scale, $\Lambda=F/S$, and
$C_i=3Y^2/5,3/4,4/3$ for fundamental representations and 0 for singlet
representations.

The mass parameters (\ref{bc1},\ref{bc2}) serve as boundary conditions
for the renormalization group equations at the messenger scale. In
order to determine the superparticle spectrum we run these parameters
from the messenger scale to the weak scale using the two-loop
renormalization group equations.  As before we impose radiative
electroweak symmetry breaking and determine the Higgs boson and
Higgsino masses $M_H$ and $|\mu|$ for a given $\tan\beta$.  Hence, the
parameter space of the model under consideration is $$\tan\beta, \quad
M, \quad\Lambda, \quad n_5, \quad n_{10}, \quad \lambda,\quad {\rm
sgn}(\mu),$$ where
$\lambda$ is the value of the messenger Yukawa coupling at the grand
unification scale (we assume a single Yukawa coupling). Note that a
physical messenger spectrum requires $M>\Lambda$.

After we determine the superpartner spectrum at the weak scale, we
apply the same weak-scale gauge coupling threshold corrections as in
the supergravity model. However, the messenger sector gives rise to
additional corrections.  Not only are the one- and two-loop
renormalization group equations altered above the messenger scale
\cite{RGEm}, but there are also new threshold corrections at the scale
$M$. A degenerate SU(5) multiplet does not affect the prediction of
$\alpha_s$ at one loop. However, the evolution of the messenger Yukawa
couplings from the grand unification scale down to the messenger scale
splits the messenger multiplets. If we decompose a 5+$\overline5$ pair
of messenger fields into their doublet and triplet components,
respectively denoted $L$ and $D$, then the messenger sector
superpotential below the grand unification scale becomes
\begin{equation}
W=\lambda_L S M_L\overline M_L + \lambda_D S M_D \overline M_D\ .
\end{equation}
Thus the messenger doublet (triplet) ends up with mass
$\lambda_LS$ ($\lambda_DS$). According to Eq.~(\ref{mas}),
this mass splitting leads to the threshold correction
(the $10+\overline{10}$ fields decompose into (SU(2),SU(3))
representations $Q$ (2,3), $U$ (1,3), and $E$ (1,1))
\footnote{There is an additional negligible correction due to the
splitting within each U(1), SU(2) or SU(3) multiplet. Each logarithm
in Eq.~(\ref{mcor}) is replaced according to $\ln(\lambda_1/\lambda_2)
\rightarrow \ln(\lambda_1/\lambda_2) +
(1/12)\ln[1-(F/\lambda_1S^2)^2]/[1-(F/\lambda_2S^2)^2]$.}
\begin{equation}
\Delta\alpha_s^{-1}(M) = {9n_5\over14\pi}\ln{\lambda_L\over\lambda_D}
~+~{n_{10}\over14\pi}\label{dam}\biggl(15\ln{\lambda_Q\over\lambda_U}
+6\ln{\lambda_Q\over\lambda_E}\biggr)\ .\label{mcor}
\end{equation}
The Yukawa couplings are evaluated at the messenger scale,
either $\lambda_LS\simeq\lambda_DS$ or $\lambda_QS\simeq
\lambda_US\simeq\lambda_ES$. Note that at one loop order
$\Delta\alpha_s^{-1}(M_Z)=\Delta\alpha_s^{-1}(M)$.

To determine the messenger Yukawa couplings at the messenger scale we
solve the renormalization group equations which are of the form
\cite{RGEm}
\begin{equation}
{d\lambda_i\over dt} = {\lambda_i\over16\pi^2}\left(
\sum_jc_j\lambda_j^2 - \sum_ka_kg_k^2\right) + \cdots
\end{equation}
where the dots indicate that there may be extra contributions due to
interactions of the singlet with the supersymmetry breaking sector
fields.  These extra contributions are the same for all the messenger
fields so they do not affect the ratios of Yukawa couplings.  Note
that for a large initial value of the messenger Yukawa coupling the
renormalization group evolution is initially dominated by the Yukawa
term which leads to the same evolution for the various messenger
fields. Hence, there will be less splitting if the initial value is
large. We illustrate this in Fig.~3 where we show the ratio of
messenger Yukawa couplings at the messenger scale vs. the starting
value at the grand unification scale.  In the 10+$\overline{10}$ case
we see that $\lambda_Q/\lambda_U$ ($\lambda_Q/\lambda_E$) varies from
1.2 to 1.3 (2.3 to 2.6).  Hence the splitting is small and confined to
a narrow range of values. Plugging this splitting into Eq.~(\ref{dam})
we find that the messenger threshold correction results in at most
$\Delta\alpha_s(M_Z)=-0.003$ for $n_{10}=1$, $+0.001$ for $n_5=1$, and
$+0.005$ for $n_5=3$.

\begin{figure}[t]
\epsfysize=2.5in
\epsffile[-75 240 25 540]{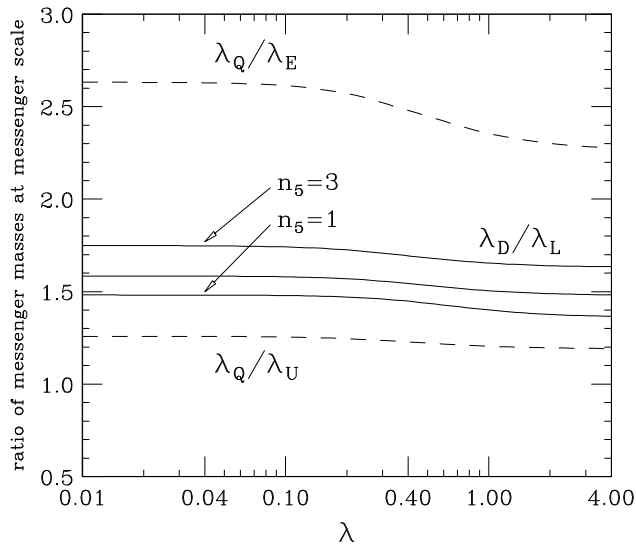}
\begin{center}
\parbox{5.5in}{
\caption[]{\small The ratios of the messenger masses at the messenger
scale vs. the value of the Yukawa coupling at the unification scale.
Both the $n_{10}=1$ (dashed) and $n_5=1,2,3$ (solid) cases and are
shown.}}
\end{center}
\end{figure}

Combining all the effects together we show the full result for the
prediction of $\alpha_s(M_Z)$ in the gauge-mediated model in the
$M/\Lambda$, $\Lambda$ plane in Fig.~4, with $\tan\beta=4$, $n_5=1$
and $\lambda=3$.  We see slightly smaller numbers than in the
supergravity case (Fig.~2). Relative to other choices of $n_5,~n_{10}$
and $\lambda$, this case results in the smallest
values of $\alpha_s(M_Z)$. As before the result is not very sensitive
to the value of $\tan\beta$. Changing the value of the messenger
Yukawa coupling does not change the value of the strong coupling
significantly.

The dashed lines in Fig.~4 indicate that the region of parameter space
with the least fine tuning ($|\mu|/M_Z\simeq2$) corresponds to the
largest value of $\alpha_s(M_Z)\simeq0.137$. The strong coupling can be
reduced significantly at the cost of fine tuning. In the fine tuned
region $|\mu|/M_Z\simeq10$, $\alpha_s(M_Z)\simeq0.125$.

\begin{figure}[t]
\epsfysize=2.5in
\epsffile[-75 240 25 540]{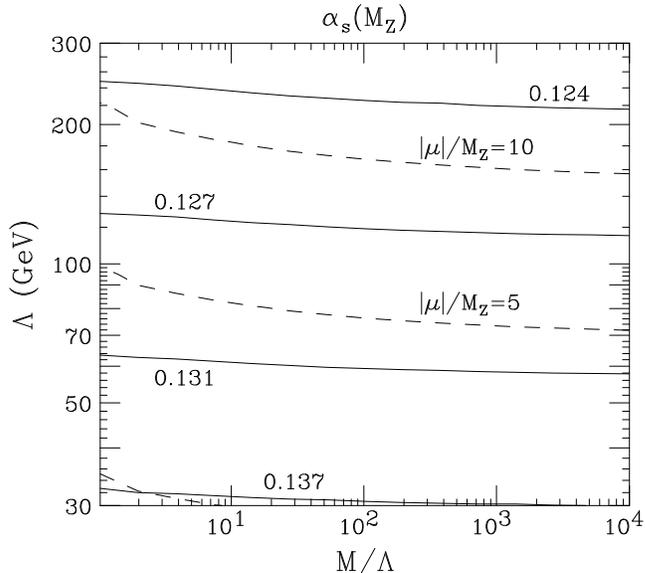}
\begin{center}
\parbox{5.5in}{
\caption[]{\small Contours of $\alpha_s(M_Z)$ in the gauge
mediated model with $n_5=1$, $\tan\beta=4$ and $\lambda=3$.
The dashed lines show contours of $|\mu|/M_Z=2,$ 5 and 10.}}
\end{center}
\end{figure}

We summarize the predicted values of $\alpha_s$ in the gauge-mediated
models in Table 1. Here we set $m_t=175$ GeV, $\tan\beta=4$, and
$\mu>0$.  We find that if $|\mu|<10M_Z$, $\alpha_s(M_Z)$ is greater
than 0.124. This is 2 standard deviations larger than the measured
value.  Hence, we are faced with the fact that, like the standard
model, supersymmetric models do not predict that the gauge couplings
meet.  However the discrepancy in the standard model case is
enormous. The small discrepancy in the MSSM can be accounted for, and
is required by, threshold corrections at the grand unification scale.

\begin{table}[t]
\begin{center}
\begin{tabular}{|c|cc|cc|}
\hline
&\multicolumn{2}{|r|}{$|\mu|<10M_Z$~~~~~~}
&\multicolumn{2}{c|}{$|\mu|<2M_Z$}
\\\hline
& $\lambda=0.01$ & $\lambda=3$ & $\lambda=0.01$ & $\lambda=3$
\\\hline
$n_5=1$ & $>\,0.125$ & $>\,0.124$ & $>\,0.137$ & $>\,0.136$
\\\hline
$n_5=3$ & $>\,0.126$ & $>\,0.125$ & $>\,0.139$ & $>\,0.136$
\\\hline
$n_{10}=1$&$>\,0.127$& $>\,0.128$ & $>\,0.137$ & $>\,0.140$
\\\hline
\end{tabular}
\parbox{5.5in}{
\caption[]{\small Summary of predictions for $\alpha_s(M_Z)$ in the
gauge-mediated models.}}
\end{center}
\end{table}

\section{\large GUT threshold corrections}

At the GUT scale there are incomplete SU(5) multiplets (most notably
the color triplet Higgs bosons) and there can be split multiplets as
well. These fields give rise to threshold corrections which are
expected to be of order a couple of per cent (or larger if
$\alpha_{\rm GUT}$ is larger).

If we take the all three gauge couplings as input at the weak scale
and run them up to the grand unification scale $M_{\rm GUT}$ (which is
defined to be the scale where $g_1$ and $g_2$ meet) we find a
discrepancy between the value of $g_3$ and $g_1=g_2$.  We define the
discrepancy $\varepsilon_g$ as
\begin{equation}
g_3(M_{\rm GUT}) = g_2(M_{\rm GUT})(1+\varepsilon_g)\ .
\end{equation}
If we fix $\alpha_s(M_Z)=0.118$ the discrepancy is negative.
We show a scatter plot of the discrepancy in Fig.~5 for three
different models: the supergravity model, the messenger model with
$n_5=1$, and the messenger model with $n_5=n_{10}=1$ \cite{scatter}.
For the supergravity model and the messenger model with $n_5=1$,
$\varepsilon_g$ varies from about $-3$ to $-1\%$ as $|\mu|$ varies
from 100 to 1000 GeV. For the messenger model with $n_5=n_{10}=1$,
$\varepsilon_g$ is larger because $\alpha_{\rm GUT}$ is larger.

\begin{figure}[t]
\epsfysize=2in
\epsffile[50 230 150 390]{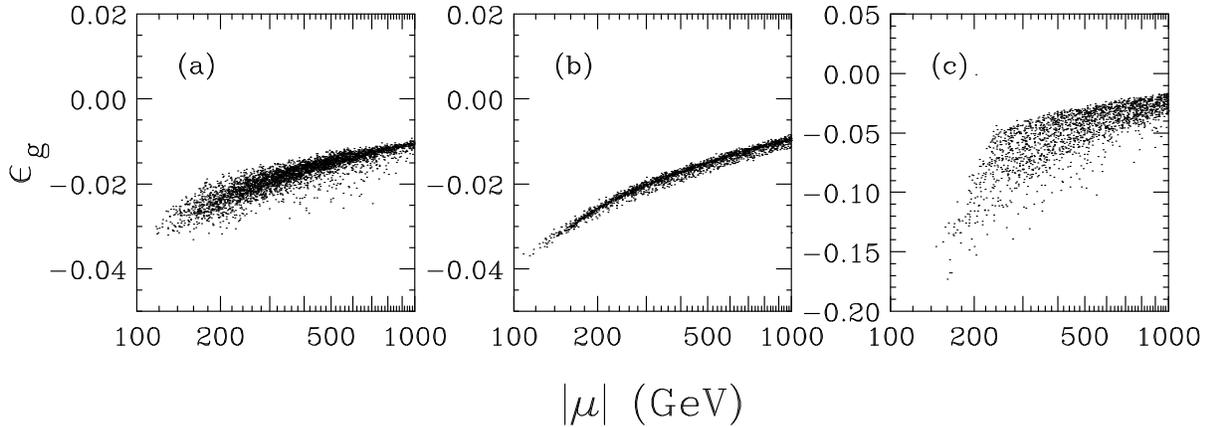}
\begin{center}
\parbox{5.5in}{
\caption[]{\small The discrepancy $\varepsilon_g$ with
$\alpha_s(M_Z)=0.118$ vs. $|\mu|$ in (a) the supergravity model, and
the messenger model with (b) $n_5=1$ and (c) $n_5=n_{10}=1$.}}
\end{center}
\end{figure}

In a grand unified theory there will be a discrepancy due to the grand
unification threshold corrections\footnote{There are possibly
additional corrections due to higher dimension operators suppressed by
$M_{\rm GUT}/M_{\rm Planck}$. See Ref.~\cite{nir} for a discussion.}.
In any particular model of grand unification we can calculate
$\varepsilon_g$ as a function of the GUT model parameter space. By
varying the parameters over their allowed range we can see whether the
model is consistent with the `measured' values of $\varepsilon_g$
shown in Fig.~5.  Here we give two examples, the minimal SU(5) model
\cite{minSU5} and the missing doublet SU(5) model \cite{MDM}.

In the minimal SU(5) model, we find the correction \cite{bmp,eg}
\begin{equation}
\varepsilon_g={3\alpha_{\rm GUT}\over 10\pi}\ln{M_{H_3}\over
M_{\rm GUT}}\ ,
\end{equation}
where $M_{H_3}$ is the triplet Higgs boson mass. It is constrained to
be larger than about 10$^{16}$ GeV by the lower limits on the nucleon
lifetimes \cite{plife}.  Both $M_{\rm GUT}$ and the bound on $M_{H_3}$
are functions of the supersymmetric parameter space. The bound on
$M_{H_3}$ is typically such that $M_{H_3}>M_{\rm GUT}$, so that in
most of the minimal SU(5) model parameter space $\varepsilon_g$ is
positive. From Fig.~5 we know that in order to be compatible with
gauge coupling unification $\varepsilon_g$ must be negative. Hence,
the minimal SU(5) model is not compatible with coupling constant
unification.

The minimal SU(5) model contains a 5+$\overline{5}$ of Higgs fields.
The doublet parts are the MSSM Higgs fields with order $M_Z$ masses.
The triplet parts mediate nucleon decay via dimension 5 operators.
In order to be compatible with the lower bound on the nucleon
lifetimes the triplet Higgs particles must have GUT scale masses.  In
general it is problematic to find a GUT model which naturally yields
light Higgs doublets and heavy triplets. In the minimal SU(5) model
this doublet-triplet splitting is imposed by fine tuning
superpotential parameters.  The missing doublet model elegantly solves
the doublet-triplet splitting problem by a judicious choice of Higgs
representations. The 75, 50 and $\overline{50}$ representations are
employed, and when the 75 gets a vev the superpotential term 75 50 5
generates a mass for the triplet, but not for the doublet.

In the missing doublet model we find \cite{bmp,egp}
\begin{equation}
\varepsilon_g = {3\alpha_{\rm GUT}\over10\pi}
\biggl\{\ln{M^{\rm eff}_{H_3}\over M_{\rm GUT}} - 9.72\biggr\}\ .
\end{equation}
We have defined an effective triplet Higgs mass $M_{H_3}^{\rm eff}$
which enters into the nucleon decay amplitude in the same way as in
the minimal SU(5) model, so the same bounds apply.  However, the
splitting in the 75-dimensional representation gives rise to a
negative correction, such that in the missing doublet model
the discrepancy $\varepsilon_g$ is negative, just as it must be in
order to be consistent with the measured values of $g_1(M_Z),
~g_2(M_Z)$ and $g_3(M_Z)$. In fact, in each of the three models shown
in Fig.~5, the allowed range of $\varepsilon_g$ in the missing doublet
model just about overlaps the required values. Hence, the missing
doublet model is consistent with gauge coupling unification.

\section{\large Conclusion}

In the end what we really do when we investigate gauge coupling
unification is constrain the physics at the grand-unification
scale. The weak-scale measurements of the gauge couplings imply that
$\varepsilon_g$ is negative. We can calculate $\varepsilon_g$ in
various grand unified models to see whether the grand unified model
parameter space can accommodate the required value. Here we showed that
this is not possible in the minimal SU(5) model, but that it is
possible in the missing doublet SU(5) model. In other words,
the missing doublet model requires that $\varepsilon_g$ is negative,
which corresponds with the measured values of the gauge couplings.
Similarly, there are SO(10) models which can accommodate the `measured'
$\varepsilon_g$, and others which cannot. Gauge coupling unification
remains an effective constraint on model building.


\begin{thebibliography}{99}
\bibitem{scatter} In the scatter plots we set $m_t=175$ GeV,
sgn$(\mu)=\pm1$, we allow $\tan\beta$ to take any value consistent
with perturbative unification and electroweak symmetry breaking, and
we vary the inputs as follows: in the supergravity model
$M_{1/2}\le500$ GeV, $M_0\le1000$ GeV and
$|A_0|\le3\sqrt{M_0^2+4M_{1/2}^2}$.  In the gauge-mediated models,
$\Lambda\le300$ TeV, $1.03\le M/\Lambda\le 10^4$, and
$0.03\le\lambda\le3$. We impose that the (first or second generation)
squark masses be greater than 220 GeV, $m_{\tilde\chi^+}>65$ GeV,
$m_{\tilde g}>170$ GeV, $m_{\tilde \ell}>45$ GeV and $m_h>60$ GeV. For
references on these bounds see \cite{pbmz}.
\end{thebibliography}
\end{document}